\documentclass[prd,aps,preprint,tightenlines,superscriptaddress,nofootinbib]{revtex4-1}
\usepackage{epsfig}
\usepackage{amsmath}

\begin{document}

\title{Gluon Distribution Functions and Higgs Boson Production at Moderate 
Transverse Momentum}
\author{Peng Sun}
\affiliation{Center for High Energy Physics, Peking University, Beijing 100871, China}
\author{Bo-Wen Xiao}
\affiliation{Department of Physics, Pennsylvania State University, University Park, PA
16802, USA}
\author{Feng Yuan}
\affiliation{Nuclear Science Division, Lawrence Berkeley National Laboratory, Berkeley,
CA 94720, USA}
\affiliation{Center for High Energy Physics, Peking University, Beijing 100871, China}
\affiliation{RIKEN BNL Research Center, Building 510A, Brookhaven National Laboratory,
Upton, NY 11973, USA}

\begin{abstract}
We investigate the gluon distribution functions and their contributions to
the Higgs boson production in $pp$ collisions in the transverse
momentum dependent factorization formalism. 
In addition to the usual azimuthal symmetric transverse momentum
dependent gluon distribution, we find that the azimuthal correlated 
gluon distribution also contributes to the Higgs boson production.
This explains recent findings on the additional contribution in 
the transverse momentum resummation for the Higgs boson production
as compared to that for electroweak boson production processes.
We further examine the small-x naive $k_t$-factorization in the dilute region
and find that the azimuthal correlated gluon distribution contribution is consistently 
taken into account, and the result agrees with the 
transverse momentum dependent factorization formalism.
We comment on the possible breakdown of the naive $k_t$-factorization in the dense medium
region, due to the unique behaviors for the gluon distributions.
\end{abstract}

\maketitle

\section{Introduction}

Recently, several studies have found that the transverse momentum resummation
for the gluon-fusion processes differ from those for the quark-antiquark annihilation (
electroweak boson/the Drell-Yan lepton pair production) processes
in the Collins-Soper-Sterman (CSS) framework~\cite{Collins:1984kg,Collins:1981uk}, in particular,
in the Higgs boson production~\cite{Catani:2010pd} and di-photon production~\cite{Nadolsky:2007ba}
processes. Similar results have been found in the context of the soft-collinear-effective
theory formalism for the Higgs boson production~\cite{Mantry:2009qz}. 
These results have raised some concern on the derivation of the CSS
formalism and the associated factorization argument for the gluon-gluon 
fusion processes. In this paper, we re-examine the transverse momentum 
dependent (TMD) factorization for Higgs boson production in $pp$ collisions. We 
find that there is an additional contribution in the leading power in the 
TMD factorization from the azimuthal correlated transverse momentum
dependent gluon distribution~\cite{werner}. 
With the complete TMD factorization results, the CSS resummation 
stands for the gluon-gluon fusion process.

Meanwhile, the azimuthal correlated gluon distribution, also referred as
the linearly polarized gluon distribution, has been recently discussed in the 
context of the transverse momentum dependent factorization formalism
in, for example, the dijet-correlation in deep inelastic scattering process~\cite{Boer:2010zf},
di-photon production in $pp$ collisions~\cite{Qiu:2011ai}.
This gluon distribution will lead to the azimuthal asymmetries in these
reactions, and the experimental observation shall provide important information
on it. Moreover, in Ref.~\cite{Metz:2011wb} it was found that the azimuthal correlated 
gluon distribution has an unique behavior at small-$x$ from the
saturation model~\cite{Gribov:1984tu,Mueller:1985wy,McLerran:1993ni}
calculations. This property emphasizes its special role in the study of the small-$x$ gluon saturation
in the high energy scattering processes, in particular, at the planed electron-ion
collider~\cite{Boer:2011fh,xiao}. In this paper, we will also examine the TMD factorization
for the Higgs boson production at small-$x$, by taking into account the 
azimuthal correlated gluon distribution contribution, and compare to the result 
obtained from the well-known small-$x$ $k_t$-factorization formalism
in the dilute region~\cite{Hautmann:2002tu,Lipatov:2005at}, 
where the two gluon distribution functions are the same at small-$x$ limit.  
The consistency between these two frameworks shed important
insights on the factorization property for the hard processes 
at small-$x$. However, in the dense medium and small
transverse momentum region, the azimuthal correlated gluon distribution is 
different from the usual gluon distribution function, and we can not 
write the Higgs boson production as the simple naive $k_t$-factorization
formalism. This indicates that the naive $k_t$-factorization breaks 
down even for the color-neutral particle production in the dense region 
in hadronic scattering processes as also found in Ref.~\cite{FillionGourdeau:2008ij}, 
similar to the situation for the heavy quark-antiquark pair production process~\cite{Blaizot:2004wv}. 

The rest of the paper is organized as follows. In Sec. II, we introduce the
leading order transverse momentum dependent gluon distributions, including the
usual one and the azimuthal correlated gluon distribution function. We will
also discuss the Collins-Soper evolution for these functions, which are important
for the transverse momentum resummation. We will show that the azimuthal correlated gluon
distribution also contributes to the Higgs boson production in $pp$ collisions.
The CSS resummation is provided in Sec.III, where the gluon distributions are
calculated in terms of the integrated parton distributions at large transverse momentum
(small $b_\perp$). In Sec.IV, we extend our discussions to the small-$x$ region,
where the TMD factorization and $k_t$-factorization formalisms are compared.
We conclude our paper in Sec. V.

\section{Transverse Momentum Dependent Gluon Distributions and Higgs Boson Production}

The transverse momentum dependent factorization is an important step to derive the CSS
resummation for the Higgs boson production in $pp$ collisions~\cite{Collins:1984kg}.
Following the Drell-Yan example, in our previous calculations~\cite{Ji:2005nu}, we have studied the 
factorization for the Higgs boson production in terms of the transverse momentum 
dependent gluon distributions, where however the azimuthal correlated gluon distribution
function was not considered. In the following, we will find that its contribution is at the same
order as the usual gluon distribution in the TMD factorization formalism. We consider,
in general case, the Higgs boson production in $pp$ collisions,
\begin{equation}
P_A+P_B\to H_0+X\ ,
\end{equation}
where $H_0$ represents the scalar-Higgs boson, and the 
hadron $A$ is moving $+\hat z$ direction and $B$ along the $-\hat z$.
Let us first introduce the spin-independent TMD gluon distributions, which can be defined
through the following matrix~\cite{Collins:1981uw,Mulders:2000sh,Ji:2005nu,Meissner:2007rx},
\begin{eqnarray}
{\cal M}^{\mu\nu}(x,k_\perp,\mu,x\zeta,\rho)&=&\int\frac{d\xi^-d^2\xi_\perp}{P^+(2\pi)^3}
    e^{-ixP^+\xi^-+i\vec{k}_\perp\cdot \vec\xi_\perp}\nonumber\\
    &&~~\times 
    \left\langle P|{F_a^{+_\mu}}(\xi^-,\xi_\perp)
{\cal L}^\dagger_{vab}(\xi^-,\xi_\perp) {\cal L}_{vbc}(0,0_\perp)
F_c^{\nu+}(0)|P \right\rangle\ ,\label{kg1}
\end{eqnarray}
where $F^{\mu\nu}_a$ is the gluon field strength tensor. The 
light-cone components are defined as $k^\pm=(k^0\pm
k^3)/\sqrt{2}$. In the above equation, $P^+=P_A^+$ represent the light-cone momentum of the
hadron $A$, and $x$ is the longitudinal momentum fraction carried by the gluon,
whereas $k_\perp$ is the transverse momentum. For the TMD parton distributions,
the gauge link ${\cal L}_v$ depends on the process~\cite{Dominguez:2010xd}. In this paper, we focus on the
Higgs boson production, and the gauge link is from the past: $ {\cal L}_v(\xi^-,{\xi_\perp};-\infty)=
P\exp\left(-ig\int_{-\infty}^{0} d\lambda v\cdot A(\lambda v + \xi) \right) $ in the
covariant gauge, where $A^\mu=A^\mu_c t^c$ is the gluon potential in the adjoint
representation, with $t^c_{ab}=-if_{abc}$.
In a singular gauge, a transverse gauge link at the spatial infinity has to be included as 
well. Four-velocity $v^\mu$
is an off-light-cone vector $v^\mu=(v^-,v^+,v_\perp=0)$ with
$v^-\gg v^+$ to regulate the light-cone singularity for the TMD
parton distributions. With this parametrization, the TMD
parton distributions will depend on $\zeta^2=(2v\cdot P)^2/v^2$. 
(In later part, we also use $\zeta_1^2$ for the TMD gluon
distribution from hadron $A$: $\zeta_1^2=(2v\cdot P_A)^2/v^2$.)
An evolution equation respect to $\zeta^2$ is called the
Collins-Soper evolution equation, and can be used
to resum the large logrithms~\cite{Collins:1984kg,Collins:1981uk}. 

In the context of the TMD factorization, we take into account the leading power
contribution in terms of $P_\perp/M$ where $P_\perp$ and $M$ are the transverse momentum
and mass for the Higgs particle. To obtain the full result in the TMD factorization,
we have to take into account the contributions from all the leading power gluon distribution functions.
The leading power expansion of the matrix ${\cal M}^{\mu\nu}$ in the unpolarized nucleon
contains two independent TMD gluon distributions~\cite{Mulders:2000sh,Meissner:2007rx,
Boer:2010zf,Qiu:2011ai},
\begin{equation}
{\cal M}^{\mu\nu}(x,k_\perp)=\frac{1}{2}\left[xg(x,k_\perp) g_\perp^{\mu\nu}
+xh_g(x,k_\perp)\left(\frac{2k_\perp^\mu k_\perp^\nu}{k_\perp^2}-g_\perp^{\mu\nu}\right) \right]\ ,\label{kg2}
\end{equation}
where $g_\perp^{\mu\nu}$ is defined as 
$g_\perp^{\mu\nu}=-g^{\mu\nu}+(P_A^\mu P_B^\nu+P_A^\nu P_B^\mu)/P_A\cdot P_B$.
In the above parameterizations, $g(x,k_\perp)$ is the usual azimuthal symmetric TMD gluon
distribution, and $h_g(x,k_\perp)$ the azimuthal correlated TMD
gluon distribution function. $h_g$ vanishes when integrating over transverse
momentum for the matrix ${\cal M}^{\mu\nu}$, which means there is no integrated
gluon distribution $h_g(x)$. In the literature, this function was also 
called ``linearly polarized" gluon distribution. However, in order
to differ from the true linearly polarized gluon distribution defined for the
spin-1 hadrons~\cite{Jaffe:1989xy,Artru:1989zv}, we prefer to use the name
of ``azimuthal correlated" gluon distribution for the spin-$1/2$ hadrons,
following the notation of Ref.~\cite{Catani:2010pd}. 
Similar functional form has been discussed in the generalized parton 
distribution for the gluons as well (see, for example, Ref.~\cite{Hoodbhoy:1998vm}).

As mentioned above, the TMD parton distribution functions
depend on the energy of the parenting hadrons, through the
variable $\zeta^2=(2v\cdot P)^2/v^2\approx 2(P^+)^2 v^-/v^+$. 
This equation is better presented in the impact parameter
space~\cite{Collins:1981uw},
\begin{equation}
\zeta\frac{\partial}{\partial\zeta}g(x,b_\perp,x\zeta,\mu,\rho)=(K_g+G_g)g(x,b_\perp,x\zeta,\mu,\rho)
\ ,\
\end{equation}
where the gluon distribution in the impact parameter space $g(x,b)$ 
is the Fourier transform of the TMD distribution: $g(x,b_\perp)=\int d^2k_\perp e^{ik_\perp\cdot b_\perp} g(x,k_\perp)$,
and $K_g$ and $G_g$ are soft and hard evolution kernels,
respectively. 
It is straightforward to extend the above equation to that for the azimuthal correlated
gluon distribution $h_g$~\cite{Idilbi:2004vb,Kang:2011mr},
\begin{equation}
\zeta\frac{\partial}{\partial\zeta}\tilde{h}_g^{\mu\nu}(x,b,x\zeta,\mu,\rho)=(K_g+G_g)\tilde h_g^{\mu\nu}(x,b,x\zeta,\mu,\rho)
\ ,\label{csh}
\end{equation}
where $\tilde h_g$ is defined as 
\begin{equation}
\tilde h_g^{\mu\nu}(x,b_\perp)=\frac{1}{2}\int {d^2k_\perp} e^{ik_\perp\cdot b_\perp}
\left(\frac{2k_\perp^\mu k_\perp^\nu}{k_\perp^2}-g_\perp^{\mu\nu}\right)h_g(x,k_\perp)\ .
\end{equation}
In particular, we find that the sum of $K_g+G_g$ are the same for the above two
equations and, at one-loop order, read,
\begin{equation}
K_g+G_g=-\frac{\alpha_sC_A}{\pi}\ln \frac{x^2\zeta^2
b^2}{4}e^{2\gamma_E-\frac{3}{2}} \ .\label{csh1}
\end{equation}
Furthermore, the $K_g$
and $G_g$ obey the following renormalization group equation,
\begin{equation}
\mu\frac{\partial K_g}{\partial \mu}=-\mu\frac{\partial
G_g}{\partial \mu}=-\gamma_{Kg}=-2\frac{\alpha_sC_A}{\pi} \ .
\end{equation}
The above evolution equations will be used to perform the 
transverse momentum resummation for the Higgs boson 
production in $pp$ collisions. 

For the TMD gluon distribution from hadron $B$, we can formulate
similarly. We will also introduce an off-light-cone vector 
$\bar v^\mu=(\bar v^-,\bar v^+,\bar v_\perp=0)$ with
$\bar v^+\gg \bar v^-$ to regulate the associated light-cone singularity,
and energy dependent variable $\zeta_2^2=4(\bar
v\cdot P_B)^2/\bar v^2$. The same Collins-Soper evolution
equations can be derived as well.

The Higgs boson production in the gluon-fusion process
can be calculated from the effective lagrangian,
\begin{equation}
{\cal L}_{eff}=-\frac{1}{4}g_\phi\Phi F^{a}_{\mu\nu}F^{a\mu\nu} \
,
\end{equation}
in the heavy top quark limit, where $\Phi$ is the scalar field and $g_\phi$ is the effective
coupling. We will use this effective lagrangian in the following calculations. We note
the finite top quark mass effects will not change our general discussions.
The leading order perturbative calculation produces Higgs particle with
zero transverse momentum. Finite transverse momentum is generated
by higher order gluon radiation contributions. However, at low transverse momentum
these high order corrections introduces large logarithms of $\ln^2(M^2/P_\perp^2)$,
which need to be resummed to make reliable predictions. The TMD 
factorization is a appropriate way to perform this resummation. 
In other words, at low transverse momentum $P_\perp\ll M$, the transverse
momentum dependence can be factorized into various leading power
TMD parton distributions, and the higher order corrections can be factorized
into the relevant hard factors. The hard factors in the TMD factorization 
does not depend on the transverse momentum, and the resummation
can be performed by solving the Collins-Soper evolution equations
for the associated TMD parton distributions. 
The lesson from the recent studies~\cite{Catani:2010pd} is that, in the TMD factorization,
we have to include all leading power contributions. In particular, the azimuthal
correlated gluon distribution $h_g$ was completely ignored in the previous study~\cite{Ji:2005nu}. 
It is straightforward to obtain this contribution in the TMD formula, and a similar
factorization can be formulated as well. After adding this
contribution, the Higgs boson production at low transverse momentum $P_\perp\ll M$
can be written as,
\begin{eqnarray}
\frac{d^3\sigma(M^2,P_\perp,y)}{d^2{P}_\perp
dy}&=&\sigma_0\int
d^2\vec{k}_{1\perp}d^2\vec{k}_{2\perp}d^2\vec{\ell}_\perp
\delta^{(2)}(\vec{k}_{1\perp}+\vec{k}_{2\perp}+\vec{\ell}_\perp-\vec{P}_\perp)\nonumber\\
&&\left\{ x_1g(x_1,k_{1\perp}) x_2g(x_2,k_{2\perp})S(\ell_\perp,\mu,\rho)H(M^2,\mu,\rho)\right.\nonumber\\
&&\left.+\left(\frac{2(k_{1\perp}\cdot k_{2\perp})^2}{k_{1\perp}^2k_{2\perp}^2}-1\right)
x_1h_g(x_1,k_{1\perp}) x_2h_g(x_2,k_{2\perp}) \right.\nonumber\\
&&~~\times \left.S_h(\ell_\perp,\mu,\rho)H_h(M^2,\mu,\rho)\right\}
\ , \label{fac1}
\end{eqnarray}
where we follow the notations of Ref.~\cite{Ji:2005nu}, $\sigma_0$ is the leading-order scalar-particle production
from two gluons, $\sigma_0={\pi g_\phi^2}/{64}$, and $y$ and
$P_\perp$ are Higgs particle's rapidity and transverse momentum, respectively.
At low-transverse momentum, the longitudinal-momentum fractions
$x_1$ and $x_2$ for the two incident gluons are related to the
scalar particle's rapidity $y$ through $x_1={Me^y/\sqrt{S}}$ and
$x_2={Me^{-y}/\sqrt{S}}$, where $S$ is the total center-of-mass
energy squared $S=(P_A+P_B)^2$. $\zeta_1$ and $\zeta_2$ are
defined above as $\zeta_1^2=4(v\cdot P_A)/v^2$ and $\zeta_2^2=4(\bar
v\cdot P_B)/\bar v^2$ and $\rho$ is a scheme-dependent parameter to
separate contributions to the soft and hard factors~\cite{Ji:2005nu}.
The above factorization result is accurate at leading power in
$P_\perp^2/M^2$ at low transverse momentum. In particular,
the interference between $g$ and $h_g$ is power suppressed 
in this limit.

The factorization for the first term with the usual gluon distribution
follows the previous argument~\cite{Ji:2005nu}, and the relevant hard
and soft factors have been calculated at one-loop order. Similar
calculations can be done for the second term in Eq.~(\ref{fac1}).
In particular, at one-loop order, we can factorize the gluon radiation
contributions to the different factors in the factorization formula
depending on the kinematic regions of the radiated gluon. 
For example, if the radiated gluon is parallel to the incoming hadron $A$,
we factorize its contribution to the TMD gluon distribution $h_g$ from
$A$. If it is parallel to the hadron $B$, we include that contribution
to the TMD gluon distribution $h_g$ from $B$. When the gluon 
momentum is soft, we factorize its contribution to the soft factor.
The hard factor is calculated from the hard gluon radiation in the virtual
diagrams, because the real gluon radiation is power suppressed if all momentum
components are hard in order of $M$. We have done the explicit calculations
to show this factorization at one-loop order~\cite{sun}. 

It is easy to find that the soft factors for the above two terms
are identical: $S_h(\ell_\perp)=S(\ell_\perp)$. This is because the soft
gluon radiations do not depend on the spin/polarization, and 
are defined identically for these two terms. In the one-loop calculations,
the hard factor are extracted from the virtual diagrams for the cross section
and parton distribution calculations, and we find that the hard 
factors are also the same, $H_h(M^2)=H(M^2)$~\cite{sun},
\begin{eqnarray}
H_h^{(1)}(M^2,\mu^2,\rho)&=&H^{(1)}(M^2,\mu^2,\rho)\nonumber\\
&=&\frac{\alpha_sC_A}{\pi}\left[\ln\frac{M^2}{\mu^2}
    \left(2\beta_0+\frac{1}{2}\ln\rho^2-\frac{3}{2}\right)-
    \frac{3}{4}\ln\rho^2+\frac{1}{8}\ln^2\rho^2+\pi^2+\frac{7}{2}\right]\
    ,\nonumber\\\label{hardh}
\end{eqnarray}
where a special coordinate system has been chosen in which
$x_1^2\zeta_1^2=x_2^2\zeta_2^2=\rho M^2$. In a recent calculation
for the single-spin dependent observable, a similar conclusion was also
found~\cite{Kang:2011mr}, which may indicate that all the hard factors in the TMD 
factorization are independent of the spin/polarization.

It is convenient to write down the TMD factorization formula in 
the impact parameter space,
\begin{eqnarray}
\frac{d^3\sigma(M^2,P_\perp,y)}{d^2P_\perp
dy}&=&\sigma_0\int\frac{d^2\vec{b}}{(2\pi)^2}e^{-iP_\perp\cdot
b_\perp}W(x_1,x_2,b,M^2) \ ,
\end{eqnarray}
where $W$ contains contribution from the two terms in the TMD
factorization,
\begin{eqnarray}
W(x_1,x_2,b,M^2)&=&W_g(x_1,x_2,b,M^2)+W_h(x_1,x_2,b,M^2) \ , \label{wb}
\end{eqnarray}
and $W_g$ and $W_h$ represent the contributions from the 
usual gluon distribution $g(x,b)$ and the azimuthal correlated 
gluon distribution $h_g(x,b)$, respectively, 
\begin{eqnarray}
W_g(x_1,x_2,b,Q^2)&=& S(b,\mu,\rho)H(M^2,\mu,\rho)
x_1g(x_1,b,\mu,\rho M^2,\rho) x_2g(x_2,b,\mu,\rho M^2,\rho) \label{fac2p} \\
W_h(x_1,x_2,b,Q^2)&=&2 S(b,\mu,\rho)H(M^2,\mu,\rho)x_1
\tilde h_g^{\mu\nu}(x_1,b,\mu,\rho M^2,\rho) x_2\tilde h_g^{\mu\nu}(x_2,b,\mu,\rho M^2,\rho)\  ,
\label{fac2}
\end{eqnarray}
where the$W_h$ comes from the specific tensor
structure in the factorization formula Eq.~(\ref{fac1}).
The convolutions in the transverse-momentum space now reduce to
products in the impact parameter $b$-space. In the factorization formula,
the large logarithms will show up as $\ln M^2b^2$ in the various 
factors in the above equations. We need to resum these large logarithms.

\section{Resummation}

The large logarithms in the factorization formulas in the last section are
resummed by following the Collins-Soper-Sterman method. The two terms
in Eq.~(\ref{wb}) satisfy the Collins-Soper evolution equation separately,
\begin{equation}
\frac{\partial W_{g,h}(x_i,b,M^2)}{\partial \ln M^2}=(K+G') W_{g,h}(x_i,b,M^2)
\ ,
\end{equation}
where $K$ and $G'$ are soft and hard evolution kernels. Since the two 
gluon distributions obey the same Collins-Soper evolution equation and
the hard factors are the same, the evolution kernels are the same as well. 
Combining the Collins-Soper evolution equations for the TMD gluon 
distributions of Eqs.~(\ref{csh},\ref{csh1}) and the 
hard factors at one-loop order of Eq.(~\ref{hardh}), we find that,
\begin{equation}
K+G'=-\frac{\alpha_sC_A}{\pi}\ln
\left(\frac{M^2b^2}{4}e^{2\gamma_E-2\beta_0}\right) \ ,
\end{equation}
where the $\rho$ dependence between various terms cancels out. 
Solving the above evolution equations, we obtain,
\cite{Collins:1984kg}
\begin{eqnarray}
W_{g,h}(x_i,b,M^2)=e^{-{\cal S}_{Sud}^{g,h}(M^2,b,C_1/C_2)}
W_{g,h}(x_i,b,C_1^2/C_2^2/b^2) \ ,
\end{eqnarray}
where the large logarithms are included in the Sudakov form factors, 
\begin{equation}
{\cal S}_{Sud}=\int_{C_1^2/b^2}^{C_2^2M^2}\frac{d
\mu^2}{\mu^2}\left[\ln\left(\frac{C_2^2M^2}{\mu^2}\right)
A(C_1,\mu)+B(C_1,C_2,\mu) \right]\ .
\end{equation}
Here $C_1$ and $C_2$ are two parameters of order one. The
functions $A$ and $B$ can be expanded perturbatively $\alpha_s$,
$A=\sum\limits_{i=1}^\infty
A^{(i)}\left(\frac{\alpha_s}{\pi}\right)^i$ and
$B=\sum\limits_{i=1}^\infty
B^{(i)}\left(\frac{\alpha_s}{\pi}\right)^i$. Because the $A$ coefficients
come from soft factor which are the same for the two terms $W_g$ and $W_h$,
we expect $A$ will be the same as well. On the other hand, $B$ coefficients come from
the hard factors in the TMD factorization formulas. 
Therefore, they could be different~\cite{deFlorian:2000pr}.
However, our one-loop calculations lead to the same hard factors and the same
$B$ coefficients for $W_g$ and $W_h$. 
We expect that the effects discussed in Ref.~\cite{deFlorian:2000pr} 
do not affect our calculations, and we conjecture that the $B$ coefficients
will be the same for these two terms at higher orders too.
With this, we can combine the above two terms together as,
\begin{eqnarray}
W(x_i,b,M^2)=e^{-{\cal S}_{Sud}(M^2,b,C_1/C_2)}\left[
W_{g}(x_i,b,C_1^2/C_2^2/b^2)+W_{h}(x_i,b,C_1^2/C_2^2/b^2) \right]\ ,\label{wb1}
\end{eqnarray}
where $S_{Sud}$ represents the universal Sudakov form factor for the 
Higgs boson production. Up to the one-loop order, we have verified this
result.

The last step of the complete CSS resummation is to formulate the $W_g$ and 
$W_h$ of the right hand side of Eq.~(\ref{wb1}) at lower scale $C_1^2/C_2^2b^2$ 
in terms of the integrated parton distributions,
\begin{eqnarray}
W_{g}(x_i,b,C_1^2/C_2^2/b^2)&=&\sum_{ij}\int \frac{dx_1'}{x_1'}\frac{dx_2'}{x_2'}x_1'f_i(x_1',\mu)x_2'f_j(x_2',\mu)\nonumber\\
&&\times
C_{g/i}(x_1/x_1',C_1/C_2/b/\mu)C_{g/j}(x_2/x_2',C_1/C_2/b/\mu) \ ,\label{wgg}\\
W_{h}(x_i,b,C_1^2/C_2^2/b^2)&=&\sum_{ij}\int \frac{dx_1'}{x_1'}\frac{dx_2'}{x_2'}x_1'f_i(x_1',\mu)x_2'f_j(x_2',\mu)\nonumber\\
&&\times
C_{h/i}(x_1/x_1',C_1/C_2/b/\mu)C_{h/j}(x_2/x_2',C_1/C_2/b/\mu) \ ,\label{whg}
\end{eqnarray}
where $f_{i,j}$ reprent the integrated quark/gluon distribution functions. For integrated gluon
distribution, there is only the usual one, whereas the counterpart of $h_g$ does not 
exist. The $C=\sum\limits_{i=0} C^{(i)}\left(\frac{\alpha_s}{\pi}\right)^i$ coefficient functions can be calculated perturbatively. The coefficients
$C_{g/i}$ have been calculated up to two-loop order~\cite{Catani:2010pd},
where $C_{h/i}$ are also calculated up to $\alpha_s$ order. In their calculations, 
the cross section $W(b)=W_g(b)+W_h(b)$ in the impact parameter ($b_\perp$) 
space is written as perturbative expansion of $\alpha_s$, from which the 
relevant coefficients are extracted by comparing with the Eqs.~(\ref{wgg},\ref{whg}).
In the following, we will show how we can calculate $C_{h/i}$ from the 
TMD factorization formula Eq.~(\ref{fac2}).

To calculate the $W_h$ in Eq.~(\ref{whg}), we compute the azimuthal
correlated gluon distribution $\tilde h_g^{\mu\nu}$ in terms of the integrated 
quark/gluon distribution functions and substitute into the factorization 
formula Eq.~(\ref{fac2}). First, we write down a similar factorization 
form for $\tilde h_g^{\mu\nu} (b_\perp)$,
\begin{equation}
\tilde h_g^{\mu\nu}(x,b_\perp)=\frac{1}{2}\left(g_\perp^{\mu\nu}-\frac{2b_\perp^\mu b_\perp^\nu}{b_\perp^2}\right)\int\frac{dx'}{x'}\tilde C_{h/i}(x/x',b_\perp,\mu) x'f_{i}(x',\mu) \ ,\label{hgexp}
\end{equation}
where the pre-factor of $\frac{1}{2}\left(g_\perp^{\mu\nu}-\frac{2b_\perp^\mu b_\perp^\nu}{b_\perp^2}\right)$
comes from the basic Lorentz structure for this function\footnote{In order to keep
this factor traceless in the dimensional regulation calculations, the factor $2$ in the bracket should be replaced by 
$d-2$ where $d=4-\epsilon$ denotes the dimension.
In our following calculations of Eqs.~(\ref{hg},\ref{hq}), since there are no divergence 
in the Fourier transform, we use $d=4$.}.
We know that there is no integrated $h_g$ gluon distribution, which immediately
leads to the zeroth order of $\alpha_s$ expansion of the above equation vanishes.
As a consequence, the zeroth order of $C_{h/i}$ in Eq.~(\ref{whg}) vanish as well,
\begin{equation}
C_{h/q}^{(0)}=C_{h/g}^{(0)}=0 \ . \label{chg}
\end{equation}
At order of $\alpha_s$, we can generate the azimuthal correlated gluon distribution
from the integrated quark/gluon distribution functions. For example, the contribution
from the integrated gluon distribution is,
\begin{equation}
h_g(x,k_\perp)=\frac{\alpha_s}{\pi^2}C_A\frac{1}{k_\perp^2}\int \frac{dx'}{x'} \frac{1-\xi}{\xi} g(x')\ ,
\end{equation}
where $\xi=x/x'$. The Fourier transform into the impact parameter space leads to,
\begin{equation}
\tilde h_g^{\mu\nu}(x,b)=\frac{1}{2}\left(g_\perp^{\mu\nu}-\frac{2b_\perp^\mu b_\perp^\nu}{b_\perp^2}\right)\frac{\alpha_s}{\pi}C_A
\int \frac{dx'}{x'} \frac{1-\xi}{\xi} g(x')\ .\label{hg}
\end{equation}
This Fourier transform does not generate any divergence, which is consistent 
with the factorization formula of Eq.~(\ref{hgexp}). Because the non-zero
leading order expansion of Eq.~(\ref{hgexp}) is at order $\alpha_s$, the right 
hand side is associated with the leading order gluon distribution, and there is
no collinear divergence. 
An interesting consequence is that the non-zero leading order coefficients do not 
depend on the factorization scale~\cite{Catani:2010pd}. However, 
from the factorization formula Eq.~(\ref{hgexp}), at order of $\alpha_s^2$,
we will find out the Fourier transform will lead to a collinear divergence which 
shall be absorbed into $\alpha_s$ order splitting of the integrated gluon distribution
function. This indicates that order $\alpha_s^2$ coefficients 
$\tilde C_{h/i}^{(2)}$ (and consequently the following $C_{h/i}^{(2)}$) 
will depend on the factorization scale. 

Similar results are obtained for the azimuthal correlated gluon distribution
in terms of the integrated quark distribution with the color-factor $C_F$ instead of $C_A$,
\begin{equation}
\tilde h_g^{\mu\nu}(x,b)=\frac{1}{2}\left(g_\perp^{\mu\nu}-\frac{2b_\perp^\mu b_\perp^\nu}{b_\perp^2}\right)\frac{\alpha_s}{\pi}C_F
\int \frac{dx'}{x'} \frac{1-\xi}{\xi} q(x')\ . \label{hq}
\end{equation}
Combining Eqs.~(\ref{hg},\ref{hq}) with Eq.~(\ref{fac2}), we obtain,
\begin{equation}
C^{(1)}_{h/q}=C_F{(1-\xi)},~~~C^{(1)}_{h/g}=C_A{(1-\xi)} \ ,
\end{equation}
which reproduces the relevant resummation formula in Ref.~\cite{Catani:2010pd}.

From the above derivation, we find that the different resummation formalism
for the gluon-gluon fusion processes as compared to that for the Drell-Yan
lepton pair production process comes from the fact that there are
two independent TMD gluon distribution functions at the leading order which contribute
to the Higgs boson production at the same order in the limit of $P_\perp\ll M$.
Although there are perturbative at different order in terms of the 
integrated quark/gluon distribution functions, we have to take into 
account the contributions from both functions in order to completely 
describe the Higgs boson production at low transverse momentum 
$P_\perp\ll M$. In particular, in certain kinematic region such as 
small-$x$ region discussed in next section, the azimuthal correlated gluon
distribution is as important as the usual one, where we have to include its
contribution.

\section{Small-$x$ $k_t$-factorzation}

The two TMD gluon distribution functions at small-$x$ have unique properties as recently 
discussed in Ref.~\cite{Metz:2011wb}. Applying these results into the
factorization formula of Eq.~(\ref{fac1}), we will be able to study the factorization
of Higgs boson production in small-$x$ region, where the well-known
$k_t$-factorization formalism~\cite{Gribov:1984tu} has been applied too.
In this section, we discuss the Higgs boson production in the small-$x$
region, and examine the so-called naive $k_t$-factorization approach in this
kinematic region.

From Eqs.~(\ref{hg}), we notice that the gluon splitting contribution to the azimuthal 
correlated gluon distribution has the same small-$x$ enhancement as the usual
gluon distribution. Therefore, we expect the similar BFKL evolution for $h_g(x,k_\perp)$
at small-x in the dilute regime~\cite{xiao,xiao1,raju,iancu}, since the operator
definition of $h_g(x,k_\perp)$ at low-$x$ is also related to the quadrupole.
As a consequence, the azimuthal correlated gluon distribution
will be as important as the azimuthal symmetric one in this kinematic limit.
In particular, from the saturation model calculations of Ref.~\cite{Metz:2011wb}, we know
that the azimuthal correlated gluon distribution function is the same as the usual gluon
distribution function at small-$x$ in the dilute region with $k_\perp\gg Q_s$, where
$Q_s$ is the characteristic scale in the saturation model. This is also consistent 
with the expectation from the BFKL evolution for these two functions~\cite{xiao}.
Therefore, in this region, we can combine the two contributions in the factorization 
formula Eq.~(\ref{fac1}) into one,
\begin{equation}
\frac{d^3\sigma}{dyd^2P_\perp}=\sigma_0\int d^2 k_{1\perp} d^2 k_{2\perp} \delta^{(2)}(P_\perp-k_{1\perp}-k_{2\perp})
x_1g(x_1,k_{1\perp}) x_2g(x_2,k_{2\perp}) \frac{2(k_{1\perp}\cdot k_{2\perp})^2}{k_{1\perp}^{2}k_{2\perp}^{2}} \ ,
\end{equation}
where we have used $g(x,k_\perp)$ to represent both $g$ and $h_g$ distribution functions,
and neglected higher order corrections from the hard and soft factors in Eq.~(\ref{fac1}). 
The above result is exactly the same as that obtained in the naive-$k_t$ 
factorization~\cite{Hautmann:2002tu,Lipatov:2005at} by taking the small
transverse momentum limit $P_\perp\ll M$~\footnote{The difference
between the results in Refs.~\cite{Hautmann:2002tu,Lipatov:2005at} vanishes in this limit.}. 
By using the proper physical gluon polarization~\cite{Gribov:1984tu}, 
one automatically takes into account the contribution from the azimuthal correlated gluon distribution 
in the naive $k_t$-factorization approach. 

However, in the dense medium (large nucleus or extremely small-$x$),
and in particular when $k_\perp\sim Q_s$, the azimuthal correlated gluon distribution is different 
from the usual gluon distribution~\cite{Metz:2011wb} if they follow the 
definitions in Eq.~(\ref{kg1},\ref{kg2}).
They are appropriate definitions for the gluon distribution in 
the Higgs boson production process~\cite{Dominguez:2010xd}. 
Therefore, we can not combine these two terms into one universal 
structure as suggested in the naive $k_t$-factorization at small-$x$. 
This indicates that the naive-$k_t$ factorization breaks down even
for the color-neutral particle production in the dense medium
in the hadronic scattering processes. Similar conclusion has also
been drawn for the $\eta'$ particle production in $pA$ collisions
in the saturation model calculations~\cite{FillionGourdeau:2008ij}. 
However, because of the large Higgs mass, we should be able to
modify the naive-$k_t$ factorization to establish an effective
$k_t$-factorization for its production at low transverse momentum
$P_\perp\ll M$, following the similar study in Ref.~\cite{Dominguez:2010xd}. 
This will lead to consistent results as the TMD factorization of Eq.~(\ref{fac1})
with the small-$x$ gluon distributions calculated in the dense region.
An explicit calculation, including high order corrections,
will be very important to investigate the QCD factorization 
property for the hard processes at small-$x$.

\section{Conclusions}

In summary, in this paper, we have investigated the transverse momentum
dependent gluon distribution functions and the Higgs boson production in $pp$
collisions in the transverse momentum dependent factorization approach. We found
that the azimuthal correlated gluon distribution contributes to the Higgs boson
production in the leading power of $P_T/M$. After taking into account this
contribution, we will be able to explain recent findings on the resummation
for the Higgs boson production at moderate transverse momentum. It will be 
interesting to extend this study to the di-photon production process
and the associated resummation formalism~\cite{Nadolsky:2007ba,Qiu:2011ai}.

We further extended our discussion to the small-$x$ region, where we compared
the TMD factorization result with the well-known naive $k_t$-factorization result,
and found that they are consistent in the dilute region. We expect they will differ
in the dense region, which may indicate the naive $k_t$-factorization is violated 
even for the neutral particle production at small-$x$ region. 

\begin{acknowledgments}
When this paper was finishing, we noticed that there were several studies~\cite{werner} on the
same topic, and all agree with each other.
We thank C.P. Yuan for bringing to our attention the results of Refs.~\cite{Catani:2010pd,Nadolsky:2007ba},
and the valuable discussions on this topic. 
We also thank L. McLerran, D. Mueller, J.W. Qiu, M. Stratmann, R. Venugopalan,
W. Vogelsang for useful conversations.
This work was supported in part by the U.S. Department of Energy under the
contracts DE-AC02-05CH11231 and DOE OJI grant No. DE - SC0002145. 
\end{acknowledgments}


\begin{thebibliography}{99}

\bibitem{Collins:1984kg}
  J.~C.~Collins, D.~E.~Soper and G.~F.~Sterman,
  Nucl.\ Phys.\  B {\bf 250}, 199 (1985).

\bibitem{Collins:1981uk}
  J.~C.~Collins and D.~E.~Soper,
  Nucl.\ Phys.\  B {\bf 193}, 381 (1981)
  [Erratum-ibid.\  B {\bf 213}, 545 (1983)]
  [Nucl.\ Phys.\  B {\bf 213}, 545 (1983)].
  
\bibitem{Catani:2010pd}
  S.~Catani and M.~Grazzini,
  Nucl.\ Phys.\  B {\bf 845}, 297 (2011)
  [arXiv:1011.3918 [hep-ph]];
  [arXiv:1106.4652 [hep-ph]].

\bibitem{Nadolsky:2007ba}
  P.~M.~Nadolsky, C.~Balazs, E.~L.~Berger and C.~P.~Yuan,
  Phys.\ Rev.\  D {\bf 76}, 013008 (2007);
  [arXiv:hep-ph/0702003].

\bibitem{Mantry:2009qz}
  S.~Mantry and F.~Petriello,
  Phys.\ Rev.\  D {\bf 81}, 093007 (2010)
  [arXiv:0911.4135 [hep-ph]]; and references therein.

\bibitem{werner}
see, also, D.~Boer, W.~ J.~ den Dunnen, C.~Pisano, M.~Schlegel, W.~Vogelsang, to appear.

\bibitem{Boer:2010zf}
  D.~Boer, S.~J.~Brodsky, P.~J.~Mulders and C.~Pisano,
  Phys.\ Rev.\ Lett.\  {\bf 106}, 132001 (2011)
  [arXiv:1011.4225 [hep-ph]].
  
\bibitem{Qiu:2011ai}
  J.~Qiu, M.~Schlegel and W.~Vogelsang,
  arXiv:1103.3861 [hep-ph].
  
\bibitem{Metz:2011wb}
  A.~Metz and J.~Zhou,
  arXiv:1105.1991 [hep-ph].
 
\bibitem{Gribov:1984tu}
  L.~V.~Gribov, E.~M.~Levin and M.~G.~Ryskin,
  Phys.\ Rept.\  {\bf 100}, 1 (1983).

\bibitem{Mueller:1985wy}
  A.~H.~Mueller and J.~w.~Qiu,
  Nucl.\ Phys.\  B {\bf 268}, 427 (1986).

  \bibitem{McLerran:1993ni}
  L.~D.~McLerran and R.~Venugopalan,
  Phys.\ Rev.\  D {\bf 49}, 2233 (1994);
  Phys.\ Rev.\  D {\bf 49}, 3352 (1994).

\bibitem{Boer:2011fh}
  D.~Boer, M.~Diehl, R.~Milner, R.~Venugopalan, W.~Vogelsang, A.~Accardi, E.~Aschenauer, M.~Burkardt {\it et al.},
  [arXiv:1108.1713 [nucl-th]].

\bibitem{xiao}
F.~Dominguez, J.~W.~Qiu, B.~Xiao, F~Yuan, to be published.

\bibitem{Hautmann:2002tu}
  F.~Hautmann,
  Phys.\ Lett.\  B {\bf 535}, 159 (2002)
  [arXiv:hep-ph/0203140].
  
\bibitem{Lipatov:2005at}
  A.~V.~Lipatov and N.~P.~Zotov,
  Eur.\ Phys.\ J.\  C {\bf 44}, 559 (2005)
  [arXiv:hep-ph/0501172].
  
\bibitem{FillionGourdeau:2008ij}
  F.~Fillion-Gourdeau, S.~Jeon,
  Phys.\ Rev.\  {\bf C79}, 025204 (2009).
  [arXiv:0808.2154 [hep-ph]].
    
\bibitem{Blaizot:2004wv}
  J.~P.~Blaizot, F.~Gelis and R.~Venugopalan,
  Nucl.\ Phys.\  A {\bf 743}, 57 (2004).

\bibitem{Ji:2005nu}
  X.~Ji, J.~P.~Ma and F.~Yuan,
  JHEP {\bf 0507}, 020 (2005)
  [arXiv:hep-ph/0503015].
  
 
\bibitem{Collins:1981uw}
  J.~C.~Collins, D.~E.~Soper,
  Nucl.\ Phys.\  {\bf B194}, 445 (1982).

  
\bibitem{Mulders:2000sh}
  P.~J.~Mulders and J.~Rodrigues,
  Phys.\ Rev.\  D {\bf 63}, 094021 (2001)
  [arXiv:hep-ph/0009343].

\bibitem{Meissner:2007rx}
  S.~Meissner, A.~Metz and K.~Goeke,
  Phys.\ Rev.\  D {\bf 76}, 034002 (2007)
  [arXiv:hep-ph/0703176].
  
\bibitem{Dominguez:2010xd} see, for example, F.~Dominguez, B.~W.~Xiao and F.~Yuan, 
Phys.\ Rev.\ Lett.\ \textbf{106}, 022301 (2011);
  F.~Dominguez, C.~Marquet, B.~W.~Xiao and F.~Yuan,
  Phys.\ Rev.\  D {\bf 83}, 105005 (2011).

\bibitem{Jaffe:1989xy}
  R.~L.~Jaffe and A.~Manohar,
  Phys.\ Lett.\  B {\bf 223}, 218 (1989).

\bibitem{Artru:1989zv}
  X.~Artru and M.~Mekhfi,
  Z.\ Phys.\  C {\bf 45}, 669 (1990).
  
\bibitem{Hoodbhoy:1998vm}
  P.~Hoodbhoy and X.~D.~Ji,
  Phys.\ Rev.\  D {\bf 58}, 054006 (1998)
  [arXiv:hep-ph/9801369];
  A.~V.~Belitsky and D.~Mueller,
  Phys.\ Lett.\  B {\bf 486}, 369 (2000)
  [arXiv:hep-ph/0005028];
  M.~Diehl,
  Eur.\ Phys.\ J.\  C {\bf 19}, 485 (2001)
  [arXiv:hep-ph/0101335].
  
\bibitem{Idilbi:2004vb}
  A.~Idilbi, X.~-d.~Ji, J.~-P.~Ma, F.~Yuan,
  Phys.\ Rev.\  {\bf D70}, 074021 (2004).
  [hep-ph/0406302].
  
\bibitem{Kang:2011mr}
  Z.~-B.~Kang, B.~-W.~Xiao, F.~Yuan,
  [arXiv:1106.0266 [hep-ph]].
    
\bibitem{sun}
P.~Sun, F.~Yuan, to be published.
    
    
\bibitem{deFlorian:2000pr}
  D.~de Florian, M.~Grazzini,
  Phys.\ Rev.\ Lett.\  {\bf 85}, 4678-4681 (2000);
  Nucl.\ Phys.\  {\bf B616}, 247-285 (2001).
  [hep-ph/0108273].

\bibitem{xiao1}
  F.~Dominguez, A.~H.~Mueller, S.~Munier, B.~-W.~Xiao,
  [arXiv:1108.1752 [hep-ph]].

\bibitem{raju}
  A.~Dumitru, J.~Jalilian-Marian, T.~Lappi, B.~Schenke, R.~Venugopalan,
  [arXiv:1108.4764 [hep-ph]].
  
\bibitem{iancu}
  E.~Iancu, D.~N.~Triantafyllopoulos,
  [arXiv:1109.0302 [hep-ph]].
\end{thebibliography}
\end{document}